\documentclass[%
aps,
prl,
 reprint,
nobibnotes,
 amsmath,amssymb,
]{revtex4-2}

\usepackage{graphicx}
\usepackage{amsfonts}
\usepackage{todonotes}
\usepackage[ngerman, english]{babel}
\usepackage{mathrsfs}
\usepackage[pdfstartview=FitH,      
            breaklinks=true,        
            bookmarksopen=true,     
            bookmarksnumbered=true,  
            hidelinks
            ]{hyperref}
            
\usepackage{letltxmacro}

\LetLtxMacro{\ORIGselectlanguage}{\selectlanguage}
\makeatletter
\DeclareRobustCommand{\selectlanguage}[1]{%
  \@ifundefined{alias@\string#1}
    {\ORIGselectlanguage{#1}}
    {\begingroup\edef\x{\endgroup
       \noexpand\ORIGselectlanguage{\@nameuse{alias@#1}}}\x}%
}
\newcommand{\definelanguagealias}[2]{%
  \@namedef{alias@#1}{#2}%
}
\makeatother

\definelanguagealias{en}{english}
\definelanguagealias{de}{ngerman}

\begin{document}

\title{A Thermodynamic Non-Linear Response Relation}
\author{Tristan Holsten}
 \email{tristan.holsten@stud.uni-goettingen.de}
\affiliation{%
 Institute for Theoretical Physics, Georg-August-Universit\"at G\"ottingen, 37077 G\"ottingen, Germany
}%
\author{Matthias Kr\"uger}%
 \email{matthias.kruger@uni-goettingen.de}
\affiliation{%
 Institute for Theoretical Physics, Georg-August-Universit\"at G\"ottingen, 37077 G\"ottingen, Germany
}%

\begin{abstract}
The fluctuation-dissipation-theorem connects equilibrium to mildly (linearly) perturbed situations in a thermodynamic manner: It involves the observable of interest and the entropy production caused by the perturbation. We derive a relation which connects responses of arbitrary order in perturbation strength to correlations of entropy production of lower order, thereby extending the fluctuation-dissipation-theorem to cases far from equilibrium in a thermodynamic way. The relation is validated and studied for a 4-state-model.
\end{abstract}

\maketitle

The theoretical footing of non-equilibrium states remains a fundamental challenge, despite important progress, e.g., given by fluctuation theorems and work relations \cite{gallavottiDynamicalEnsemblesNonequilibrium1995,jarzynskiNonequilibriumEqualityFree1997,crooksEntropyProductionFluctuation1999, Seifert_2012}. When aiming at the non-equilibrium responses, the powerful \textit{fluctuation-dissipation-theorem} (FDT) provides the leading order. It states that measurements of thermodynamic fluctuations in the unperturbed equilibrium system predict behaviours of the perturbed one in linear order \cite{einsteinUberMolekularkinetischenTheorie1905,CallenWeltonIrreversibility, greenMarkoffRandomProcesses1954,kuboStatisticalMechanicalTheoryIrreversible1957,kuboFluctuationdissipationTheorem1966}. 
 With stronger perturbations, it fails, making it necessary to consider \textit{non-linear responses}, which have been addressed by the response formula from Kawasaki \cite{KawasakiNonlinearEffects}, transient time correlation functions \cite{evansStatisticalMechanicsNonequilibrium2008,petravicNonlinearResponseTimeDependent1998, fuchsIntegrationTransientsBrownian2005}, equilibrium correlations deduced for specific systems \cite{Oppenheim_1989,Bouchaud_2005, LippielloNonlinearResponse,Lucarini_2012,Diezemann_2012} as well as by generic approaches using operator formalisms \cite{kuboStatisticalMechanicalTheoryIrreversible1957, Andrieux_2007} and path integrals \cite{Colangeli_2011, basuFreneticAspectsSecond2015,basuExtrapolationNonequilibriumCoarseGrained2018, mullerCoarsegrainedSecondOrder2020}.

An important observation is that the non-linear responses are of fundamentally different nature as compared to FDT: Already at second order, dynamical details of the system enter \cite{basuFreneticAspectsSecond2015}. Determining and measuring theses dynamical details in the considered system, such as (interaction-)potentials, hinders application of the mentioned approaches to macroscopic systems with many degrees of freedom \cite{Maes_2020_Trajectories}. The thermodynamic nature is thus lost in higher orders, causing the notion of a theorem to stop at first order. 

We derive a formula for non-linear responses to arbitrary order, which is distinct from known relations by its thermodynamic appearance. 
It consists of correlations of the excess entropy and the observable of interest. These are taken out of equilibrium as well, but one order lower than the response. For the linear response, one order lower is equilibrium, so that the derived relation extends the FDT to far from equilibrium scenarios in a natural manner. We demonstrate this formula for a non-Markovian jump process by calculating the responses for time independent as well as time dependent perturbation up to third order. We also investigate  statistical convergence for the examples provided. 

Consider a classical system in weak contact with a thermal bath, described by a phase space $x_s$ at time $s$. It is in equilibrium until time $s=0$ and then driven out of equilibrium by a Hamiltonian
\begin{align}
H(x_s, s)=\varepsilon h(s) V(x_s).\label{eq:H}
\end{align}
$V(x_s)$ is a potential, $\varepsilon$ is a dimensionless expansion parameter setting the strength of the perturbation and $h(s)$ is the perturbation protocol of order unity, being finite for $  0 \leq s \leq t$. Using perturbations of potential type eases the presentation, and we will comment on other types later in the manuscript.

We continue by reviewing the standard approach via path integrals \cite{basuFreneticAspectsSecond2015,  baiesiFluctuationsResponseNonequilibrium2009,wynantsStructuresNonequilibriumFluctuations2010, Colangeli_2011} to arrive at Eq.~\eqref{eq: intermediate eq} below. 
The path weight $P^\varepsilon(\omega)$ for a path $\omega=(x_s)$, which describes the phase space configurations $x_s$ of the system on the time interval $[0,t]$, enables the computation of the expectation value $\langle O(x_t)\rangle^\varepsilon$ of a state observable $O(x_t)$ via the path integral
\begin{align}
\langle O(x_t)\rangle^\varepsilon=\int \mathcal{D}\omega O(x_t)P^\varepsilon(\omega).
\label{eq: forward expectation value}
\end{align} 
$P^\varepsilon(\omega)=e^{-A_\varepsilon(\omega)}P(\omega)$ is split into a part comprising the perturbation via the non-equilibrium action $A_\varepsilon(\omega)$   \cite{basuFreneticAspectsSecond2015, wynantsStructuresNonequilibriumFluctuations2010, baiesiFluctuationsResponseNonequilibrium2009, Colangeli_2011} and the equilibrium path weight $P(\omega)$. The latter satisfies time-reversibility $P(\omega)=P(\varTheta \omega)$. The sequence $\varTheta \omega =(\pi x_{t-s}), \, 0\leq s \leq t$, represents the time-reversed path where the operator $\pi$ reverses the sign of kinematic components. Out of equilibrium, time-reversal symmetry is broken and $P^\varepsilon(\omega)$ does not equal its time-reversed counterpart. When speaking of time-reversal, also the protocol $h(s)$ is reversed \cite{wynantsStructuresNonequilibriumFluctuations2010, mullerCoarsegrainedSecondOrder2020}, and the corresponding path weights are denoted by tilde superscripts,  $\tilde{P}^\varepsilon(\varTheta \omega)= e^{-\tilde{A}_\varepsilon(\varTheta \omega)}P(\varTheta \omega)$. 
The breaking of time-reversibility is quantified by the excess entropy flux $S_\varepsilon(\omega)$ towards the environment \cite{basuFreneticAspectsSecond2015, maesTimeSymmetricFluctuationsNonequilibrium2006, wynantsStructuresNonequilibriumFluctuations2010, maesEntropyProductionCase2007, Colangeli_2011, baiesiFluctuationsResponseNonequilibrium2009} 
\begin{align}
S_\varepsilon(\omega)=&\tilde{A}_{\varepsilon}(\varTheta\omega)-A_{\varepsilon}(\omega)\nonumber \\
=&\varepsilon\beta \left[h(t)V(x_t)-h(0)V(x_0)-\int_0^t\dot{h}(s)V(x_s)\mathrm{d}s\right],
\label{eq: entropy general}
\end{align}
where $\beta=(k_bT)^{-1}$ with $T$ the temperature and $k_B$ the Boltzmann constant. The thermodynamic role of the entropy flux $S_\varepsilon(\omega)$ becomes clear by the explicit form given in the second line: It contains no system-specific information, and can be written down without specifying the system under consideration. 

This  time-antisymmetric part of $A_{\varepsilon}(\omega)$ is complemented by the time-symmetric part $D_\varepsilon(\omega)=\frac{1}{2}[\tilde{A}_{\varepsilon}(\varTheta\omega)+A_{\varepsilon}(\omega)]$ \cite{basuFreneticAspectsSecond2015, maesTimeSymmetricFluctuationsNonequilibrium2006, wynantsStructuresNonequilibriumFluctuations2010, maesEntropyProductionCase2007, Colangeli_2011, baiesiFluctuationsResponseNonequilibrium2009}. In contrast to $S_\varepsilon(\omega)$, $D_\varepsilon(\omega)$ has no thermodynamic interpretation. A general form such as the lower line of Eq.~\eqref{eq: entropy general} is not known, but has been given for specific systems \cite{Colangeli_2011, basuFreneticAspectsSecond2015, wynantsStructuresNonequilibriumFluctuations2010}. 
To proceed, we assume that $A_\varepsilon(\omega)$ and $\tilde{A}_{\varepsilon}(\varTheta\omega)$ can be expanded around $\varepsilon=0$, i.e., $D_\varepsilon(\omega)=\varepsilon D^\prime(\omega)+\varepsilon^2D^{\prime\prime}(\omega)/2+\dots$ and $S_\varepsilon(\omega)=\varepsilon S^\prime(\omega)$, compare Eq.~\eqref{eq: entropy general} \cite{basuFreneticAspectsSecond2015}. It has  proven useful to define the expectation value of $O(x_t)$ under time-reversed dynamics \cite{mullerCoarsegrainedSecondOrder2020, basuFreneticAspectsSecond2015}
\begin{align}
\langle O(x_{t})\varTheta\rangle^\varepsilon&=\int O(x_{t})\tilde{P}^{\varepsilon}(\varTheta \omega)\mathcal{D}\omega = \langle O(x_t)\rangle=\langle O(x)\rangle,
\label{eq: backwards expectation value}
\end{align}
where $\langle O(x)\rangle=\int O(x)P(\omega)\mathcal{D}\omega$ is the equilibrium expectation value of the state observable. The equality in Eq.~\eqref{eq: backwards expectation value} can be derived via causality and time-reversal relations \cite{basuFreneticAspectsSecond2015, mullerCoarsegrainedSecondOrder2020}. Subtracting Eq.~\eqref{eq: backwards expectation value} from Eq.~\eqref{eq: forward expectation value} and expanding in powers of $\varepsilon$ gives
\begin{align}
\langle O(x_{t})\rangle^\varepsilon =\langle O(x) \rangle+ \nonumber
& \sum_{n=0}^\infty \frac{\varepsilon^n}{n!} \left\langle \frac{\mathrm{d}^n}{\mathrm{d} \varepsilon^n}\left( e^{-D_\varepsilon(\omega) +\frac{\varepsilon}{2}S^\prime(\omega) }\right.\right.\\
&\left.\left.-e^{-D_\varepsilon(\omega) -\frac{\varepsilon}{2} S^\prime(\omega) }\right)\bigg \vert_{\varepsilon=0}O(x_{t})\right\rangle.
\label{eq: intermediate eq}
\end{align}
Executing the derivatives and expanding Eq.~\eqref{eq: intermediate eq} in $\varepsilon$ results in the known expression of the responses via equilibrium correlations \cite{basuFreneticAspectsSecond2015, Colangeli_2011}. In linear order of $\varepsilon$, $D_\varepsilon(\omega)$ drops out, yielding the FDT. As mentioned, already the second order term involves $D_\varepsilon(\omega)$, as do higher orders, implying that non-linear responses are not of thermodynamic nature; while this statement is of principal interest, it makes determination of the non-linear responses challenging and laborious, particularly for complex systems \cite{basuFreneticAspectsSecond2015, mullerCoarsegrainedSecondOrder2020, heldenMeasurementSecondorderResponse2016, Colangeli_2011, Maes_2020_Trajectories}. To make progress, we use an identity based on the Leibniz rule \cite{bronsteinSpringerTaschenbuchMathematik2013},
\begin{align}
&\frac{\mathrm{d}^n}{\mathrm{d} \varepsilon^n} e^{-D_\varepsilon(\omega)\pm \frac{\varepsilon}{2} S^\prime(\omega)}\bigg \vert_{\varepsilon=0}\nonumber \\
=&\sum_{i=0}^n  {{n}\choose{i}}\frac{\mathrm{d}^{n-i}}{\mathrm{d} \varepsilon^{n-i}}e^{-D_\varepsilon(\omega)}\bigg \vert_{\varepsilon=0} \left(\pm \frac{S^\prime(\omega)}{2}\right)^{i},
\label{eq.: product rule}
\end{align}
with the binomial coefficient ${{n}\choose{i}}$.
Plugging the identity of Eq.~\eqref{eq.: product rule} into Eq.~\eqref{eq: intermediate eq}, substituting $e^{-D_\varepsilon(\omega)}=e^{-A_\varepsilon(\omega)-\frac{\varepsilon}{2}S^\prime (\omega)}$ and using the identity once more (with $D_\varepsilon(\omega)$ replaced by $A_\varepsilon(\omega)$) yields
\begin{align}
& \langle O(x_{t})\rangle^\varepsilon=\langle O(x)\rangle +2\sum_{n=1}^{\infty} \frac{1}{n!} \varepsilon^n\sum_{i=1,\,odd}^{n} {{n}\choose{i}} \sum_{j=0}^{n-i} {{n-i}\choose{j}}\nonumber \\[1pt]
& \times (-1)^j\frac{\mathrm{d}^{n-i-j}}{\mathrm{d} \varepsilon^{n-i-j}} \left\langle \left(\frac{S^\prime(\omega)}{2}\right)^{i+j}O(x_{t})\right\rangle^\varepsilon\Bigg\vert_{\varepsilon=0}.
\label{eq: series expansion generalised formula new}
\end{align}
As required, the sum contains no  term $n=0$. 

Eq.~\eqref{eq: series expansion generalised formula new}, hereafter referred to as \textit{thermodynamic response relation} (TRR), is the main result of this manuscript. It relates the mean of the observable far from equilibrium on the left hand side to correlation functions involving excess entropy on the right hand side. Its thermodynamic meaning is thus displayed with Eq.~\eqref{eq: entropy general}. 
It is insightful to regard the first orders in $\varepsilon$ from Eq.~\eqref{eq: series expansion generalised formula new} explicitly, mirroring the responses $\chi_n$ of order $n$ ($\langle O(x_{t})\rangle^\varepsilon=\langle O(x)\rangle + \chi_1 \varepsilon+ \chi_2 \varepsilon^2\dots $),
\begin{align}
\chi_1^{TRR}(t)=&\langle S^\prime(\omega) O(x_t)\rangle \nonumber \\
\chi_2^{TRR}(t)=&\frac{\mathrm{d}}{\mathrm{d}\varepsilon}\langle S^\prime(\omega) O(x_t)\rangle^\varepsilon \bigg \vert_{\varepsilon=0}-\frac{1}{2}\langle S^\prime(\omega)^2O(x_t)\rangle\nonumber \\
\chi_3^{TRR}(t)=&\frac{1}{2}\frac{\mathrm{d^2}}{\mathrm{d}\varepsilon^2}\langle S^\prime(\omega) O(x_t)\rangle^\varepsilon\bigg \vert_{\varepsilon=0}\nonumber \\&-\frac{1}{2}\frac{\mathrm{d}}{\mathrm{d}\varepsilon}\langle S^\prime(\omega)^2O(x_t)\rangle^\varepsilon \bigg \vert_{\varepsilon=0}+\frac{1}{6}\langle S^\prime(\omega)^3O(x_t)\rangle.
\label{eq: responses new}
\end{align}
The linear response, i.e., the first line in Eq.~\eqref{eq: responses new}, resembles FDT. Higher order responses contain higher order correlations of $S^\prime(\omega)$, and also  derivatives of these correlations with respect to $\varepsilon$, so that the right hand side of Eq.~\eqref{eq: series expansion generalised formula new} requires measurements under applied perturbation. Notably, the $n$th order response is related to correlation functions up to order $n-1$. We thus interpret the TRR in Eq.~\eqref{eq: series expansion generalised formula new} as follows: Even far from equilibrium, thermodynamics allows to predict one order in perturbation strength -- This new insight contains FDT as a special case, connecting equilibrium to linear order.  

\begin{figure}[t]
  \centering
 \includegraphics[width=0.75\linewidth]{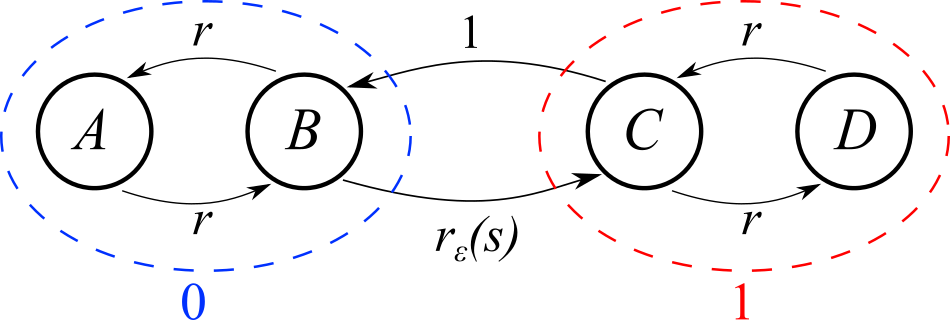}
  \caption{Sketch of the 4-state-model. The rates $r$ within the macrostates $0$ and $1$ are chosen small ($r=0.1$), so that the resulting two state model is strongly non-Markovian \cite{basuExtrapolationNonequilibriumCoarseGrained2018}. The transition rate $r_\varepsilon(s)$ equals unity  for $s<0$ and follows a perturbation protocol for $s\geq 0$.}
  \label{fig. four state model}
\end{figure}

 We illustrate and examine the TRR of Eq.~\eqref{eq: series expansion generalised formula new} in a simple model, where four states $A,B,C,D$ are connected via dimensionless transition rates as depicted in Fig.~\ref{fig. four state model}. The system's dynamics follows a simple master equation.  For illustration purposes, we pretend to be blind to  microscopic details by coarse graining: Let states $A$ and $B$ form macrostate $X=0$, and $C$ and $D$ macrostate $X=1$ \cite{basuExtrapolationNonequilibriumCoarseGrained2018}. The resulting two state system is Markovian for large values of $r$ and non-Markovian if $r$ is small. We choose $r=0.1$ to achieve the latter. This model thus allows to discuss the case of hidden degrees of freedom. As it can also be solved analytically, it makes a good test case for our purposes \cite{basuExtrapolationNonequilibriumCoarseGrained2018,mullerCoarsegrainedSecondOrder2020}. 
 
 Treating the so obtained setup via response theory is challenging, due to the presence of the mentioned hidden degrees of freedom: Eq.~\eqref{eq: intermediate eq} is not applicable, as evaluation of $D_\varepsilon(\omega)$ requires microscopic resolution \cite{wynantsStructuresNonequilibriumFluctuations2010}. This example allows to illustrate the thermodynamic type of Eq.~\eqref{eq: series expansion generalised formula new}: It is applicable despite presence of hidden degrees of freedom.
 
The system is in equilibrium for $s<0$ and perturbed for $s\geq0$ by making the transition rate which connects macrostate $0$ to macrostate $1$ a function of the dimensionless time~$s$ via the protocol $h(s)$, i.e., $r_\varepsilon(s)=e^{\varepsilon h(s)}$. This perturbation corresponds to Eq.~\eqref{eq:H} with a dimensionless potential $V(X_s)$ with  $V(0)=0$ and $V(1)=1$ \cite{basuExtrapolationNonequilibriumCoarseGrained2018, mullerCoarsegrainedSecondOrder2020}. The entropy production $S_\epsilon(\omega)$ in Eq.~\eqref{eq: entropy general} depends on $X_s$, so that Eq.~\eqref{eq: series expansion generalised formula new} acts in the space of  macrostates, and its thermodynamic nature is apparent. Once on that level, the 'internal' nature of the macrostates is not relevant concerning validity of Eq.~\eqref{eq: series expansion generalised formula new}, as required from a thermodynamic relation. 

\begin{figure}[]
\setlength{\abovecaptionskip}{-2pt}
  \centering
 \includegraphics[width=1.\linewidth]{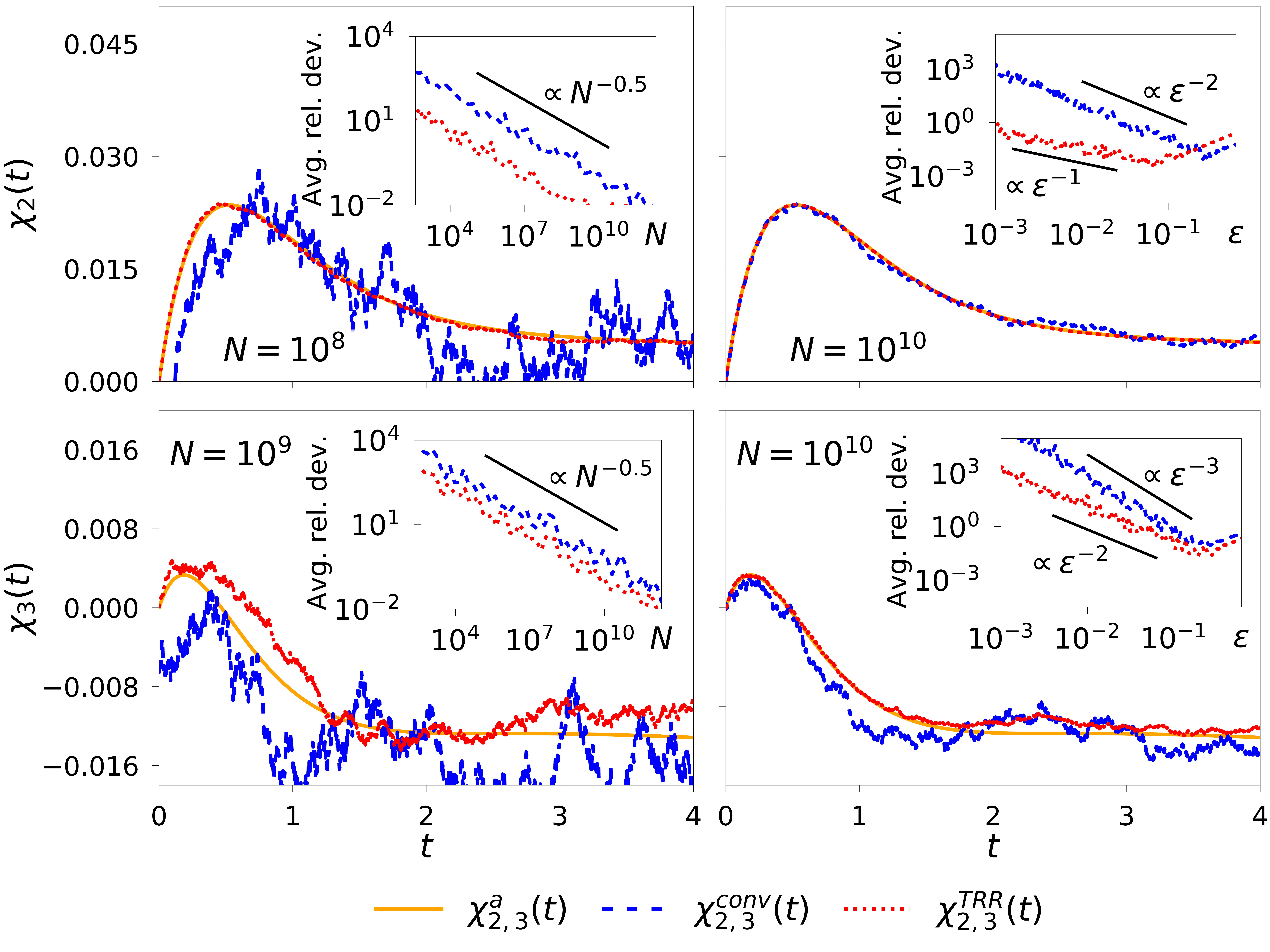}
 \caption{Second (top) and third (bottom) order responses of the 4 state model for the TIP, as functions of time $t$, found analytically or via simulations using Eq.~\eqref{eq: series expansion generalised formula new} (TRR), or the conventional method. The number of independent measurements is denoted $N$. Insets depict the relative statistical error as a function of $N$ (left) for $\epsilon=0.1$ and as a function of $\epsilon$ (right) for $N=10^{10}$.}
\label{fig:convergence test}
\end{figure}

We start with a time independent perturbation (TIP), i.e.,  $h(s)=1, s \geq 0$, so that Eq.~\eqref{eq: entropy general} simplifies to $S^\prime(\omega)=V(X_{t})- V(X_0)$ ($\beta=1$ here and in the following). 
Furthermore, we choose $O(X_t)=X_t$. We evaluate the responses up to third order, using computer simulations of the master equation, and applying Eq.~\eqref{eq: responses new}. The derivatives with respect to $\varepsilon$ appearing in Eq.~\eqref{eq: responses new} are calculated via central differencing, i.e., 
$\frac{\mathrm{d}}{\mathrm{d}\varepsilon }\langle ... \rangle^\varepsilon \vert_{\varepsilon=0}={\lim_{\varepsilon\rightarrow 0}}\frac{\langle ...\rangle^{\varepsilon}-\langle ...\rangle^{-\varepsilon}}{2\varepsilon}$ for first order, $\frac{\mathrm{d}^2}{\mathrm{d}\varepsilon^2 }\langle ... \rangle^\varepsilon \vert_{\varepsilon=0}={\lim_{\varepsilon\rightarrow 0}}\frac{\langle ...\rangle^{\varepsilon}-2\langle ...\rangle+\langle ...\rangle^{-\varepsilon}}{\varepsilon^2}$ for second order and so on  \footnote{The third order is given by:\\ $\frac{\mathrm{d}^3}{\mathrm{d}\varepsilon^3}{\langle ... \rangle^\varepsilon \vert_{\varepsilon=0}} ={\lim_{\varepsilon \rightarrow 0}} \frac{\langle ...\rangle^{2\varepsilon}-2\langle ...\rangle^{\varepsilon}+2\langle ...\rangle^{-\varepsilon}-\langle ...\rangle^{-2\varepsilon} }{2\varepsilon^3}$}. This requires a choice of $\varepsilon$. 
The results for $\chi_2^{TRR}$ and $\chi_3^{TRR}$ are shown in Fig.~\ref{fig:convergence test} as red data points using $\varepsilon=0.1$ and a simulation time step of $\Delta t=0.001$. In that curve, we also show the exact results $\chi_{2}^{a}$, $\chi_{3}^{a}$, which can be found analytically \footnote{See Supplemental Material for more details regarding the analytical expressions in the coarse-grained 4-state-model}. $\chi_2$ raises to a maximum at around $t\approx 0.5$, and then goes to zero for $t\to \infty$, for reasons of symmetry. This final decay is slow due to the internal rate $r$ being small. $\chi_3$ exhibits a similar behavior, but levels off to a finite static response at $t\to \infty$. Fig.~\ref{fig:convergence test} demonstrates the validity of Eq.~\eqref{eq: series expansion generalised formula new} for the given model, and that model also allows to test practical aspects, such as the convergence of  Eq.~\eqref{eq: series expansion generalised formula new} in the statistical sense. For $\chi_2$ and $\chi_3$, we provide two panels each with different number $N$ of independent computer 'measurements'. While the main graphs supply a qualitative impression, insets of the left panels give the  \textit{averaged relative deviation}, i.e.,  the time average of the ratio $ \vert [\chi_n(t)-\chi_n^a(t)]/\chi_n^a(t)\vert$.
The panels show the decay of this relative error with $1/\sqrt{N}$, as expected. The insets in the right panels display this relative error for a fixed $N=10^{10}$, as a function of $\varepsilon$ used in the mentioned central differencing. The curves reveal a minimum: For large values of $\varepsilon$, Eq.~\eqref{eq: series expansion generalised formula new} acquires a systematic error, while, for $\varepsilon\to 0$ and a fixed $N$, the statistical error diverges. 

The statistical quality of results obtained via Eq.~\eqref{eq: series expansion generalised formula new} can be compared to the outcome of \textit{conventional response} measurements, for which the responses $\chi_n^{conv}$ of $n$th order are obtained from perturbed data as \cite{heldenMeasurementSecondorderResponse2016, basuExtrapolationNonequilibriumCoarseGrained2018}
\begin{align}
\chi_n^{conv}(t)=\frac{1}{n!}\frac{\mathrm{d}^n}{\mathrm{d}\varepsilon^n }\langle X_{t}\rangle^\varepsilon \bigg \vert_{\varepsilon=0}.
\label{eq: responses conv}
\end{align}
One difference between evaluating Eqs.~\eqref{eq: series expansion generalised formula new} and \eqref{eq: responses conv} is already evident from the method of central differencing: The higher the order $n$, the more different experimental or simulation setups have to be used (e.g. two, i.e., $+\varepsilon$ and $-\varepsilon$ for $\chi_1^{conv}$).
The TRR, Eq.~\eqref{eq: series expansion generalised formula new} thus, for odd orders, requires a smaller number of different setups (1 versus 2 for $\chi_1$ (FDT),  2 versus 2 for $\chi_2$, 3 versus 4 for $\chi_3$). 

The results of the conventional method are displayed in Fig.~\ref{fig:convergence test} as blue data points. When comparing the two formalisms, we note similar behavior of the relative error (insets), but the scaling for $\varepsilon\to 0$ is different, with power laws that are advantageous for the TRR. The smaller the value of $\varepsilon$, i.e., the smaller a systematic error is sought, the better is the TRR in comparison to the conventional method.

\begin{figure}[t]
\setlength{\abovecaptionskip}{-2pt}
  \centering
 \includegraphics[width=1.\linewidth]{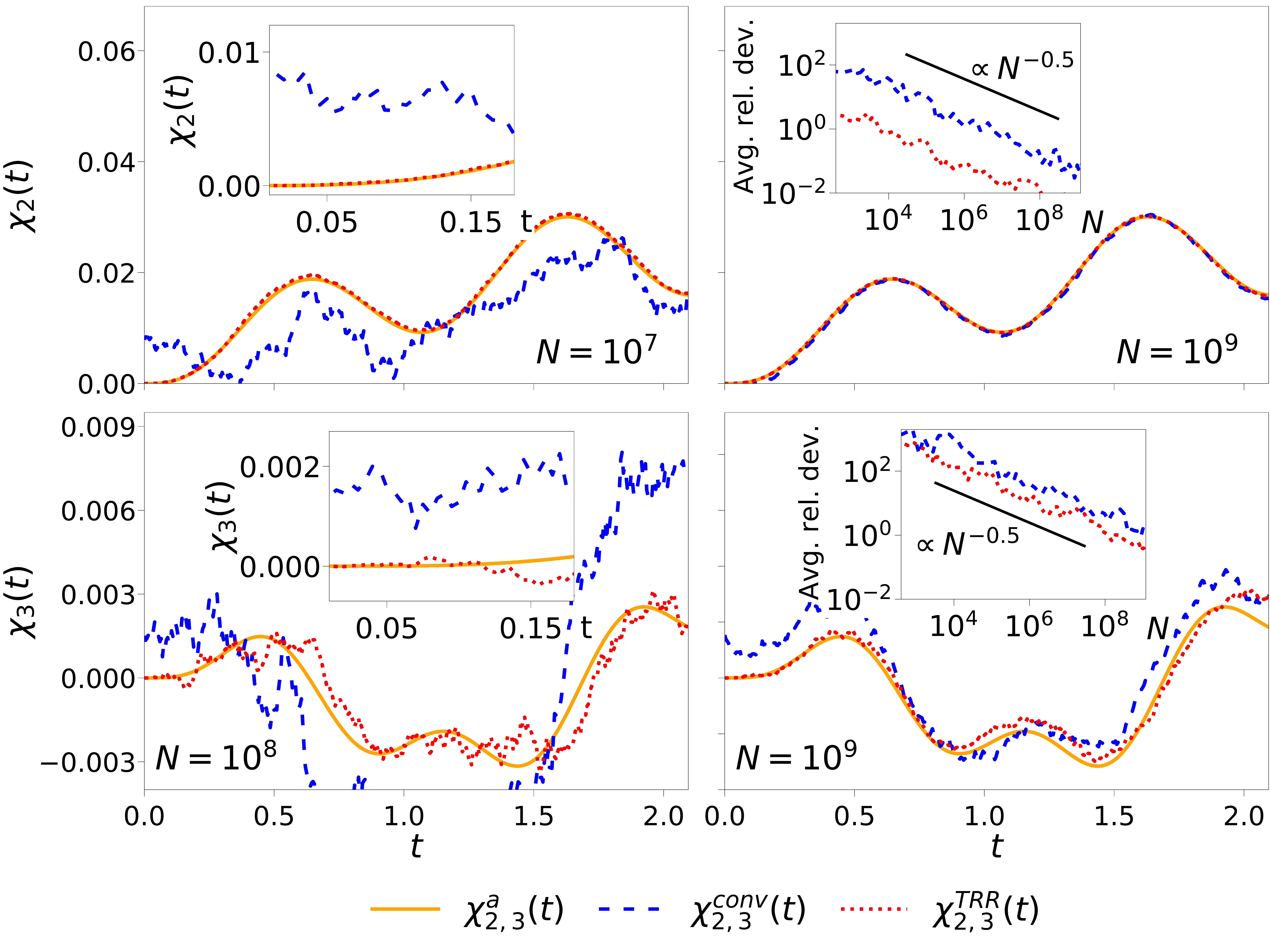}
 \caption{Second (top) and third (bottom) order responses for a sinusoidal perturbation as functions of time $t$. Responses are computed analytically, or found by simulations via Eq.~\eqref{eq: series expansion generalised formula new} (TRR) or via the conventional method. $N$ gives the number of independent measurements. Left insets present a closer view of the short time behavior. Right insets show the relative error as a function of $N$ for $\varepsilon=0.2$.}
\label{fig:convergence test_2}
\end{figure} 

We continue by investigation of a time dependent perturbation (TDP) of the form $h(s)=\sin(\omega_0 s)$, where  $\omega_0=\pi$ is chosen. Analogously to the above, second and third order responses are calculated, analytically \footnotemark[\value{footnote}], via the TRR in Eq.~\eqref{eq: responses new} and the conventional method in Eq.~\eqref{eq: responses conv}, here using $\Delta t=0.005$ and $\varepsilon=0.2$ for simulations. Fig.~\ref{fig:convergence test_2} presents the resulting curves, again demonstrating the validity of Eq.~\eqref{eq: series expansion generalised formula new}. Furthermore, as for TIP, we show the relative error as a function of $N$ for the two methods. Both scale as $1/\sqrt{N}$ as before, with the error of TRR being again smaller for the given value of $\varepsilon$.
We also show a close up view of the short time behavior, which discloses a fundamental difference. For $t\to 0$, the responses vanish. While the error of the conventional method stays finite for any $t$, the relative error diverges for $t\to0$. For this reason, we excluded times $t\leq 0.1$ when computing the time averaged relative deviation depicted in the other insets. In contrast, the TRR appears to be much more precise at short times, because $S
^\prime(\omega)$ naturally approaches zero leading to vanishing $\chi_n^{TRR}$. While qualitatively advantageous for short times, one may expect the TRR to deviate stronger for very large $t$ (larger than shown here), as the error may add up in the time integrals in $S^\prime(\omega)$. This will be investigated in future work.

A thermodynamic non-linear response relation connects responses to time correlation functions at arbitrary order of perturbation strength. With it, thermodynamics allows to predict one order in perturbation strength, a statement, which encloses the FDT at lowest order. Although we restricted the derivation of Eq.~\eqref{eq: responses new} to perturbations by a potential,  Eq.~\eqref{eq: series expansion generalised formula new} is also applicable for non-conservative perturbation forces, as long as $S_\varepsilon$ is linear in $\varepsilon$, which is a typical case.
Analyzing this relation for a coarse-grained 4-state-model displays similarities but also fundamental differences in the scaling of the statistical error compared to the conventional method. Especially for small value of perturbation strength, the new formula converges faster compared to the conventional method. 
Future work will address possible extension to perturbations around non-equilibrium steady states.

We thank Urna Basu, Laurent Helden, Gregor Ibbeken and Fenna M\"uller for useful discussions.

\end{document}


\title{A Thermodynamic Non-Linear Response Relation - Supplemental Material}
\author{Tristan Holsten}
\affiliation{%
 Institute for Theoretical Physics, Georg-August-Universit\"at G\"ottingen, 37077 G\"ottingen, Germany
}
\author{Matthias Kr\"uger}%
\affiliation{%
 Institute for Theoretical Physics, Georg-August-Universit\"at G\"ottingen, 37077 G\"ottingen, Germany
}%
\maketitle

We detail in this supplemental material how the responses of the coarse-grained 4-state-model to the time independent perturbation and to the sinusoidal perturbation are analytically calculated and give the exact expressions for the second and third order response for these perturbations.\\
The time evolution of the probability distributions of the four states $\rho_i(t)$, $i \in \{A,B,C,D\}$  is given by the master equation involving the generator matrix \cite{basuExtrapolationNonequilibriumCoarseGrained2018, mullerCoarsegrainedSecondOrder2020}
\begin{align}
M=\begin{pmatrix}                                
-r & r & 0& 0 \\                                               
r & -r-r_{\epsilon}(s)& r_{\epsilon}(s) & 0 \\ 
0& 1 &-r-1 & r\\                                         
0 & 0 & r& -r \\                                               
\end{pmatrix},
\label{eq: generator matrix}
\end{align}
The initial condition is that the system is in equilibrium at $t=0$ which corresponds to the distribution $\rho_i(0)=1/4$, $i \in \{A,B,C,D\}$. The expectation value of the observable $O(X_t)=X_t$ on the coarse-grained level is  computed by $\langle X_t\rangle^\epsilon=\rho_C(t)+\rho_D(t)$ \cite{basuExtrapolationNonequilibriumCoarseGrained2018}.

The master equation can be explicitly solved for the time independent perturbation. The second and third order response responses can be calculated by the derivatives in Eq.~(8) in the main text (for brevity, we provide all explicit forms for $r=0.1$)
\begin{align}
\chi^a_2(t)=&\frac{1}{81608}e^{-1.1 t} \big[\sqrt{101} (100-909 t) \sinh \big(\sqrt{1.01} t\big)\nonumber \\
&+9191 t \cosh \big(\sqrt{1.01} t\big)\big]
\end{align}
and
\begin{align}
\chi^a_3(t)=&\frac{1}{197817792}e^{-(\sqrt{1.01}+1.1) t} \bm{\big(}-303 t \big[\big(1819 \sqrt{101}\nonumber \\
&+18281\big) t-4 \big(507 \sqrt{101}+5093\big)\big]-4121204 \nonumber \\
&\times e^{(\sqrt{1.01}+1.1) t}+e^{2\sqrt{1.01} t} \big\{303 t \big[\big(1819 \sqrt{101}\nonumber \\
&-18281\big) t-2028 \sqrt{101}+20372\big]+21302 \sqrt{101}\nonumber \\
&+2060602\big\}-21302 \sqrt{101}+2060602\bm{\big)}.
\end{align}

For the sinusoidal perturbation, the master equation cannot be explicitly solved. Therefore, the responses are determined by expanding the formal solution, which is a time-ordered exponential, in terms of $\epsilon$ resulting in a Dyson-series \cite{mullerCoarsegrainedSecondOrder2020}. When executing the obtained integrals, the following expressions for the second and third order are obtained,
\begin{widetext}
\begin{align}
\chi_2^{a}(t)=&-\big[1616 \big(101+25 \pi ^2\big) \big(1+111 \pi ^2+25 \pi ^4\big) \big(1+444 \pi ^2+400 \pi ^4\big)\big]^{-1}e^{-(3 \sqrt{1.01}+1.1) t} \bm{\big(}-101 \pi\big(101+25 \pi ^2\big) \nonumber \\
&\times e^{(3 \sqrt{1.01}+1.1) t} \big[\big(1+178 \pi ^2+8500 \pi ^4-5000 \pi ^6\big) \sin (2 \pi  t)-3 \pi  \big(6+675 \pi ^2+5000 \pi ^4\big) \cos (2 \pi  t)+\pi  \big(6+25 \pi ^2\big) \nonumber \\
&\times \big(1+444 \pi ^2+400 \pi ^4\big)\big]+2 \sqrt{101} e^{3 \sqrt{1.01} t} \sinh \big(\sqrt{1.01} t\big) \big\{\big(1+444 \pi ^2+400 \pi ^4\big) \big[275 \pi  \big(1+5 \pi ^2\big) \sin (\pi  t)+2 \big(202\nonumber \\
&+23975 \pi ^2+6250 \pi ^4\big) \cos (\pi  t)\big]+2 \big(1+111 \pi ^2+25 \pi ^4\big) \big(-202-94194 \pi ^2-52625 \pi ^4+2500 \pi ^6\big)\big\}+202 e^{3 \sqrt{1.01} t} \nonumber \\
&\times \cosh \big(\sqrt{1.01} t \big) \big\{2 \big(1+111 \pi ^2+25 \pi ^4\big) \big(23+9756 \pi ^2+8625 \pi ^4+2500 \pi ^6\big)-\big(1+444 \pi ^2+400 \pi ^4\big) \big[25 \pi  \big(5 \pi ^2\nonumber \\
&-1\big)\sin (\pi  t)+2 \big(23+2400 \pi ^2+625 \pi ^4\big) \cos (\pi  t)\big]\big\}\bm{\big)}
\end{align}
and
\begin{align}
\chi_3^{a}(t)=&\big[1958592 \pi  \big(101+25 \pi ^2\big)^2 \big(101+100 \pi ^2\big) \big(1+111 \pi ^2+25 \pi ^4\big)^2 \big(1+444 \pi ^2+400 \pi ^4\big) \big(1+999 \pi ^2+2025 \pi ^4\big)\big]^{-1}\nonumber \\
&\times e^{-(3 \sqrt{1.01}+2.2) t} \Big(-303 e^{(3 \sqrt{1.01}+1.1) t}\big(1+444 \pi ^2+400 \pi ^4\big) \cosh \big(\sqrt{1.01} t\big)  \bm{\big\{}-101\bm{\big[}25 \pi ^8 \bm{\big(}625 \pi ^4 \big\{50 \pi ^2 \big[75 \pi ^2 \nonumber \\
&\times \big(26407+2400 \pi ^2\big)+12370777\big]+2900081621\big\}+4621856501725 \pi ^2+4476537956287 \bm{\big)}+33729449681951 \pi ^6\nonumber \\ 
&+619396582987 \pi ^4+3215953221 \pi ^2+2631858\ \bm{\big]}+2 \pi ^2 \big(101+25 \pi ^2\big) \big(101+100 \pi ^2\big) \big(1+111 \pi ^2+25 \pi ^4\big) \nonumber \\
&\times \big(1+999 \pi ^2+2025 \pi ^4\big) \big(4366+484550 \pi ^2+119375 \pi ^4\big) t-101 \big(101+25 \pi ^2\big) \big(1+111 \pi ^2+25 \pi ^4\big) \big(1+999 \pi ^2\nonumber \\\displaybreak
&+2025 \pi ^4\big) \big\{\big[25 \pi ^2 \big(38929+48940 \pi ^2+9750 \pi ^4\big)+8686\big] \cos (2 \pi  t)+\pi  \big[25 \pi ^2\big(21576+25575 \pi ^2+5000 \pi ^4\big)\nonumber \\
&+4816\big]\sin (2 \pi  t)\big\}\bm{\big\}}+101 \big(101+100 \pi ^2\big) \bm{\big[}909 e^{(3 \sqrt{1.01}+2.2) t} \pi ^2 \big(1+111 \pi ^2+25 \pi ^4\big) \big[25 \pi ^6 \big(-185551-497900 \pi ^2\nonumber \\
&+60000 \pi ^4\big)-150379 \pi ^4-1209 \pi ^2-2\big] \cos (3 \pi  t) \big(101+25 \pi ^2\big)^2-101 e^{(3 \sqrt{1.01}+2.2) t} \pi  \big(1+111 \pi ^2+25 \pi ^4\big) \big[50 \pi ^6 \nonumber \\
& \times\big(-453291-761675 \pi ^2+1395000 \pi ^4\big)-301283 \pi ^4-1011 \pi ^2-1\big] \sin (3 \pi  t) \big(101+25 \pi ^2\big)^2-3 \big(1+999 \pi ^2\nonumber \\
&+2025 \pi ^4\big) \big\{-101 e^{(3 \sqrt{1.01}+2.2) t} \big(6+3403 \pi ^2+320165 \pi ^4+1919125 \pi ^6+3312500 \pi ^8\big) \big(101 \pi+25 \pi ^3\big)^2\nonumber \\
&-2 e^{ (4 \sqrt{1.01}+1.1) t} \big(1+111 \pi ^2+25 \pi ^4\big)^2\big[125000 \pi ^8 \big(9 \sqrt{101}-101\big)+202 \big(432 \sqrt{101}-4343\big)+2500 \pi ^6 \big(4718 \sqrt{101}\nonumber \\
&-48177\big)+707 \pi ^2 \big(54819 \sqrt{101}-550981\big)+100 \pi ^4 \big(381602 \sqrt{101}-3840323\big)\big]+2 e^{(2 \sqrt{1.01}+1.1) t} \big(1+111 \pi ^2\nonumber \\
&+25 \pi ^4\big)^2 \big[125000 \pi ^8 \big(9 \sqrt{101}+101\big)+202 \big(432 \sqrt{101}+4343\big)+2500 \pi ^6 \big(4718 \sqrt{101}+48177\big)+707 \pi ^2 \nonumber \\
&\times \big(54819 \sqrt{101}+550981\big)+100 \pi ^4 \big(381602 \sqrt{101}+3840323\big)\big]\big\} \cos (\pi  t)-3 \big(\pi +999 \pi ^3+2025 \pi ^5\big)\nonumber \\
&\times \bm{\big(}-101 e^{(3 \sqrt{1.01}+2.2) t}\big\{50 \pi ^6 \big[625 \pi ^2 \big(-107+24 \pi ^2\big)-51011\big]-77699 \pi ^4-607 \pi ^2-1\big\} \big(101+25 \pi ^2\big)^2\nonumber \\
&-50 e^{(4 \sqrt{1.01}+1.1) t} \big(1+111 \pi ^2+25 \pi ^4\big)^2 \big[101 \big(\sqrt{101}-9\big)+10000 \pi ^6 \sqrt{101}+50 \pi ^4 \big(739 \sqrt{101}-4949\big)+18 \pi ^2 \nonumber \\
&\times \big(2399 \sqrt{101}-23331\big)\big]+50 e^{(2 \sqrt{1.01}+1.1) t} \big(1+111 \pi ^2+25 \pi ^4\big)^2\big[10000 \pi ^6 \sqrt{101}+101 \big(\sqrt{101}+9\big)+50 \pi ^4 \nonumber \\
&\times \big(739 \sqrt{101}+4949\big)+18 \pi ^2 \big(2399 \sqrt{101}+23331\big)\big]\bm{\big)} \sin (\pi  t)\bm{\big]}+3 e^{ (3 \sqrt{1.01}+1.1) t} \big(1+444 \pi ^2+400 \pi ^4\big) \bm{\big\{}\pi ^2 \nonumber \\
&\times \bm{\big[}-25 \pi ^6 \bm{\big(}625 \pi ^4 \big\{50 \pi ^2 \big[25 \pi ^2 \big(84528979+7435800 \pi ^2\big)+13601411543\big]+3173994538649\big\}\nonumber \\
&+4973852941113775\pi^2+4751067447679093\bm{\big)}-35480872475621649 \pi ^4-639468614663713 \pi ^2+202 \big(101\nonumber \\
&+25 \pi ^2\big) \big(101+100 \pi ^2\big) \big(1+111 \pi ^2+25 \pi ^4\big)\big(1+999 \pi ^2+2025 \pi ^4\big) \big(434+48200 \pi ^2+11875 \pi ^4\big) t\nonumber \\
&-3263596534879\bm{\big]}-101 \big(101+25 \pi ^2\big) \big(1+111 \pi ^2+25 \pi ^4\big) \big(1+999 \pi ^2+2025 \pi ^4\big) \big\{\big[25 \pi ^2 \big(391061+488540 \pi ^2\nonumber \\
&+97250 \pi ^4\big)+87264\big] \cos (2 \pi  t)+\pi  \big[25 \pi ^2 \big(218854+256375 \pi ^2+50000 \pi ^4\big)+48884\big]\nonumber \\
&\times \sin (2 \pi  t)\big\}-2670540192\bm{\big\}}  \sqrt{101}\sinh \big(\sqrt{1.01} t\big)\Big)
\end{align}
\end{widetext}

%
